# Macrosteps dynamics and the growth of crystals and epitaxial layers

Stanislaw Krukowski*, Konrad Sakowski, Paweł Strak, Paweł Kempisty, Jacek Piechota, and Izabella Grzegory

*Institute of High Pressure Physics, Polish Academy of Sciences, Sokolowska 29/37, 01-142 Warsaw, Poland*

Step pattern stability of the vicinal surfaces during growth was analyzed using various surface kinetics models. It was shown that standard analysis of the vicinal surfaces provides no indication on the possible step coalescence and therefore could not be used to elucidate macrostep creation during growth. A scenario of the instability, leading go macrostep creation, was based on the dynamics of the step train, i.e. the step structure consisting of the high (train) and low (inter-train) density of the steps. The critical is step motion at the rear of the train which potentially leads to the step coalescence i.e. creation of the double and multiple step. The result of the analysis shows that the decisive factor for the step coalescence is the step density ratio in and out of the train. The ratio lower than 2 prevents double step formation irrespective of the kinetics. For higher ratio the coalesce depends on step kinetics: fast incorporation from lower terrace stabilizes the single steps, fast incorporation from upper leads to step coalescence. The double step is slower than the single steps, so the single steps behind catch up creating multistep and finally macrostep structure. The final surface structure consists of the macrosteps and superterraces, i.e. relatively flat vicinal segments. The macrostep alimentation from lower superterrace leads to emission of the single steps which move forward. Thus the single step motion is dominant crystal growth mode in the presence of the macrosteps. These steps finally are absorbed by the next macrostep. The absorption and emission of single steps sustain the macrostep existence, i.e. the macrostep fate is determined the single step dynamics. The condition for single step emission was derived. In addition, the macrosteps are prone to creation of the overhangs which results from surface dynamics coupling to impingement from the mother phase. The angular preferential access of the bulk material to the macrostep edge, leads to the overhang instability and creation of inclusions and dislocations.

This research was funded in part by National Science Centre, Poland under grant number 2021/41/B/ST5/02764. For the purpose of Open Access, the author has applied a CC-BY-NC-ND public copyright license to any Author Accepted Manuscript (AAM) version arising from this submission.

This research was funded in part by National Science Centre, Poland under grant number 2021/41/B/ST5/02764. For the purpose of Open Access, the author has applied a CC-BY-NC-ND public copyright license to any Author Accepted Manuscript (AAM) version arising from this submission.

## I. Introduction

Growth of bulk crystals and epitaxial layers is controlled by the same step motion rules. Nevertheless, microscopic picture of the bulk and epitaxial growth differs considerably. Epitaxial growth of nanostructures requires atomic step precision as such is required to obtain quantum structures, such as quantum wells, wires or dots for optoelectronic and electronic applications [1-3]. Accordingly, epitaxial growers stride to enforce atomic step flow mode of the growth which is obtained using very low supersaturation, careful surface preparation and consequently, the slow and ordered step motion and the resulting growth rates [4,5]. Since, the total thickness of the device structure is of order of 1 ÷ 2 microns, such low rates are acceptable [6-8].

The focal point in bulk crystal growth is located differently. Bulk crystals should have uniform properties and be devoid or at least have limited concentration of the extended defects such as dislocations, disclinations, foreign phase inclusions, stacking faults, etc. [9-11]. In addition, the presence of the uncontrolled dopants, mostly foreign atoms should be minimized to the level not harmful to the planned applications [12,13]. This is accompanied by general requirement of the optimal macroscopic shape, which is enforced by optimal growth rates along the specified crystallographic directions [14,15]. Generally the growth rate of the bulk crystals should be much higher, at least two orders of magnitude higher than the growth rates of epitaxial layers. In consequence, the atomic layer control of the growing surface is of secondary importance and it is intentionally sacrificed to attain high growth rates. In consequence, majority of bulk crystal surfaces is composed of macrosteps [16-29]. Astonishingly, step bunching and creation of macrosteps is also observed in some equilibrium crystal shapes [30-33]. Nevertheless, creation of macrosteps in equilibrium is extremely rare while in growth it is universal phenomenon. Therefore, the dynamic factors are crucial in creation of macrosteps during growth. It is not still clear however which factor is primarily responsible for dynamic creation of the macrosteps.

Creation of the macrosteps in equilibrium was considered in many works. Relatively early, Cabrera and coworkers analyzed equilibrium properties of crystalline surfaces [34,35]. They showed that macrostep can stabilize the equilibrium shape for specific angular dependence of surface free energy [34]. The same line of investigations was used by Williams

This research was funded in part by National Science Centre, Poland under grant number 2021/41/B/ST5/02764. For the purpose of Open Access, the author has applied a CC-BY-NC-ND public copyright license to any Author Accepted Manuscript (AAM) version arising from this submission.

et al. who discovered presence of macrosteps for specific orientations of Si surfaces [31,32]. More extended approach was used in Monte Carlo (MC) investigations by Akutsu that lead to discovery of faceted macrosteps [36,37]. Similar step bunching effects were obtained by the introduction of the inter-ledge attraction that destabilizes regular step trains [33,36,38].

Macrostep creation during crystal growth is different phenomenon, related to surface dynamics. As such it should be investigated in conjunction with the standard step-flow model. Theoretically, crystal growth by atomic step flow mode was formulated relatively early by Burton, Cabrera and Frank (BCF) [39]. The growth proceeds via parallel step motion, nourished by surface diffusion over terraces. The steps have tendency to maintain equal spacing due to mutual diffusional repulsion [40-43]. The diffusive interaction depends on the incorporation of the surface adatoms at the steps and at the terraces [43]. The simplest case, known as BCF model assumes no supersaturation at the steps, which entails no kinetic barriers and consequently no step asymmetry [39]. Introduction of the incorporation barriers at the step opens the possibility of incorporation of step asymmetry, i.e. existence of Ehrlich-Schwoebel effect [44-47]. The effect was discovered by Ehrlich and Hudda in their investigations of surface diffusion of W adatoms at tungsten surfaces by ion-field microscopy [44]. Subsequently the effect was analyzed theoretically by Schwoebel et al., indicating that the step spacing depends on the difference in probability of incorporation of the adatoms into the steps from upper and lower terraces [46]. The effect was then shown to induce step bunching under sublimation [46].

Generally Ehrlich-Schwoebel effect is invoked as a possible force behind the two step approaching each other that may lead to bunching instability. In case of the absence of the strong asymmetry, it is argued that the step bunch expands so that vicinal surface is stable [47]. An additional approach is based on kinematic wave evolution into the shock wave that could be modeled by a simple equation on the step density [48,49]. The nonlinear version of the equation provides solitary wave solution in which the correlation between the steepness of the shock wave and the soliton speed. The different size waves move with different velocity that could therefore coalesce [50]. The usefulness of this approach is not clear as the Burgers equation is not directly related to dynamics of the diffusing adsorbate. An additional approach proposed by Chernov, assumed that the step bunching occurs due to coupling between step motion and the flow in the mother phase [51,52]. As it will be shown bulk diffusion coupling

This research was funded in part by National Science Centre, Poland under grant number 2021/41/B/ST5/02764. For the purpose of Open Access, the author has applied a CC-BY-NC-ND public copyright license to any Author Accepted Manuscript (AAM) version arising from this submission.

to surface step dynamics is evidently important factor inducing emergence of the macrostep via creation of the step trains. Therefore this early identification was confirmed by the analysis given in the present paper.

In the recent years a number of attempts were made to enrich basic BCF and Schwoebel models by introduction of additional factors that affects the step motion and leads to creation of the macrosteps during growth [53-56]. Vlachos and Jensen added effect of step disorder caused by incorporation of impurities to claim that results in bunching and finally, creation of macrosteps [53] Akutsu used a direct extension of equilibrium approach to discontinuity of surface tension angular dependence to analyze its influence on the step-step interaction [54]. Strangely enough, the macrostep size was found to decrease with the driving force increase. The height of the faceted macrostep in function of the driving force, i.e. supersaturation $\Delta\mu$ [55]. Several regimes was found, in function of the driving force increase: (i) step detachment from macrostep are inhibited, (ii) step detachment by heterogenous nucleation at the macrostep, (iii) successive step detachment are possible, (iv) macrostep vanishes and the surface is roughened kinetically [45]. Further investigations was directed to the relation between macrostep intensity and growth velocity at the vicinal surfaces [56]. It was found that the growth velocity decreases for higher faceted macrostep height.

The present paper does not follow the above mentioned routes. Macrostep creation is all-present phenomenon, observed in crystal growth from solution [16], growth from the eutectic alloys [17], vapor growth of SiC [18], molecular beam epitaxy growth of GaAs layers [21], protein lysozyme growth from mother liquor [22] or copper electrodeposition [23], growth of AlGaN layers [24,25], zinc cadmium thiocyanate crystals [26], epitaxy and solution growth of GaAlAs layers [27,28] or molecular beam epitaxy of cadmium mercury layers [29]. Thus macrostep creation is observed for all types of the crystals, grown by all types of the growth techniques. Therefore macrostep creation is not related to specific type of steps interaction, diffusive barrier anisotropy, etc. The process results from the core of crystal growth i.e. the step dynamics. The appropriate approach is to use very basic model, such as BCE model with its possible extension and to find the scenario in which emergence of the macrostep is possible. Therefore the present paper is devoted to determination of the macrostep properties at very basic level. It will start with the condition of stability of a single step train, with the analysis of the step formation and finally, the description of the basic instabilities. The analytical approach

This research was funded in part by National Science Centre, Poland under grant number 2021/41/B/ST5/02764. For the purpose of Open Access, the author has applied a CC-BY-NC-ND public copyright license to any Author Accepted Manuscript (AAM) version arising from this submission.

will be supplemented by numerical simulations of the dynamics to determine not only the approximate results but also the precise data that can be further used in more advanced analysis in the future.

The obtained results will formulate the basics of macrostep dynamics. It will be shown that macrostep formation is dynamic phenomenon related to surface dynamics at the specific conditions. The creation scenario will be supplemented by the explanation of the role of the macrostep during growth by analysis of the processes which may lead to the macrostep height increase or the decrease until final disappearance. Finally, the scenario of the possible role of macrostep in the creation of extended defects, such as inclusions or dislocations will be elucidated.

## II. The model

Fundamental equation of the surface diffusion, obtained by Burton, Cabrera and Frank (BCF) in seminal paper on crystal growth, determines the time evolution of the supersaturation $\sigma$ at the atomically flat crystal terrace [39]:

$$\left(\tau_{sur} \frac{\partial}{\partial t} - \frac{l_{sur}^2}{4} \Delta\right) \sigma = \sigma - \sigma_v \quad (1a)$$

where $\sigma_v$ is supersaturation in the vapor phase at contact with the terrace. The surface residence time of adsorbed species $\tau_{sur}$ is:

$$\tau_{sur} = \tau_o \, exp\left(\frac{\phi}{kT}\right) = \nu^{-1} \, exp\left(\frac{\phi}{k_B T}\right) \quad (1c)$$

and the (2-d) surface diffusion length $l_{sur}$ is given by

$$l_{sur}^2 = 4D\tau_{sur} = \frac{a}{2} \, exp\left\{\frac{\phi - E_{bar}}{k_B T}\right\} \quad (1b)$$

In the above formulae, $\phi$ and $E_{bar}$ are energy needed for desorption (equal to bond energy $\phi$) and energy barrier for the jump between two equivalent surface sites, respectively. The other quantities, determining the length and time scales are: $a$ and $\nu$ ($\nu = \tau_o^{-1}$), the lattice constant and the attempt frequency, respectively.

The growth occurs at the step, or more precisely, at the step kinks, the repeatable growth sources [57,58,39]. It is assumed frequently that the thermally excited kinks densely populate the steps transforming the steps into linear geometry growth pattern. The growth proceeds by step motion, building the atomic layer of the crystal body. The simplest possible assumption made by BCF was that the incorporation at the step is extremely fast, sufficient to attain local

This research was funded in part by National Science Centre, Poland under grant number 2021/41/B/ST5/02764. For the purpose of Open Access, the author has applied a CC-BY-NC-ND public copyright license to any Author Accepted Manuscript (AAM) version arising from this submission.

equilibrium. Thus the supersaturation vanishes at the step ($\sigma = 0$). Neglecting the step motion, BCF considered parallel array of straight steps separated by the distance $y$. The stationary solution of Eq. 1a is symmetric with respect of the terrace center located at $z = 0$:

$$\sigma = \sigma_v \left[1 - \frac{\cosh\left(\frac{2z}{l_{sur}}\right)}{\cosh\left(\frac{y}{l_{sur}}\right)}\right] \quad -\frac{y}{2} < z < \frac{y}{2} \tag{2}$$

where the steps are located at $z = \pm\frac{y}{2}$. The solution is periodic with the period equal the interstep distance, i.e. $y$. The diffusional flux of the atoms toward the step from the both (left and right) terraces is obtained as:

$$\vec{j}_{dif} = -2D_{sur}\nabla n_{sur} = -2D_{sur}\eta c_{sur-eq}\nabla\sigma \tag{3a}$$

where the density of surface adatoms is $n_{sur} = \eta\, c_{sur} = \eta\, c_{sur-eq}\sigma$. $c_{sur}$ and $c_{sur-eq}$ are the occupations of the surface sites at any time and in equilibrium ($0 \leq c_{sur} \leq 1$). The density of surface sites is denoted by $\eta$, in the case of the square lattice: $\eta = a^{-2}$. Therefore the atomic flux to the step is:

$$j_{dif} = \frac{l_{sur}\sigma_v c_{sur-eq}}{a^2 \tau_{sur}} \tanh\left(\frac{y}{l_{sur}}\right) \tag{3b}$$

This could be used to obtain the flux at single site:

$$R = j_{dif}a = \frac{l_{sur}\sigma_v c_{sur-eq}}{a\tau_{sur}} \tanh\left(\frac{y}{l_{sur}}\right) \tag{3c}$$

and the step velocity

$$v = Ra = \frac{l_{sur}\sigma_v c_{sur-eq}}{\tau_{sur}} \tanh\left(\frac{y}{l_{sur}}\right) = v_\infty \tanh\left(\frac{y}{l_{sur}}\right) \tag{3d}$$

where $v_\infty = \frac{l_{sur}\sigma_v c_{sur-eq}}{\tau_{sur}}$ is the step velocity for infinite distance between steps.

BCF assumption does not describe barrier controlled atom incorporation at the step. This could be done by introduction of kinetic coefficient describing the growth unit (adatom) incorporation at the step during the time interval $\tau_o$ denoted by $k$. According to theory of chemical kinetic processes the kinetic coefficient $k$ is equal to transition probability, given by Arrhenius relation $k = exp\left(-\frac{E_{bar-step}}{k_B T}\right)$. The energy barrier for the incorporation at the step site is $E_{bar-step}$. Therefore the effective incorporation of the adatoms $R$ at the step site is [59]:

$$R = kv(c_{sur} - c_{sur-eq}) = \frac{k\, c_{sur-eq}\sigma}{\tau_o} \tag{4}$$

The introduction of the kinetics of the incorporation changes the supersaturation at the terrace from zero to nonzero value. In consequence, the solution given by Eq. 2 is changed to:

This research was funded in part by National Science Centre, Poland under grant number 2021/41/B/ST5/02764. For the purpose of Open Access, the author has applied a CC-BY-NC-ND public copyright license to any Author Accepted Manuscript (AAM) version arising from this submission.

$$\sigma = \sigma_v \left[1 - \frac{\alpha \cosh\left(\frac{2z}{l_{sur}}\right)}{\cosh\left(\frac{y}{l_{sur}}\right)}\right] \quad -\frac{y}{2} < z < \frac{y}{2} \tag{5}$$

The solution coherence condition assumes the equality of the diffusive flux at the step length $a$ and incorporation flux at single site:

$$R = j_{dif}\, a \tag{6a}$$

that can be translated into the condition for $\alpha$ as

$$g\,(1-\alpha) = \alpha \tanh\left(\frac{y}{l_{sur}}\right) \tag{6b}$$

where the factor $g = \frac{2k\tau_{sur}a}{l_{sur}\tau_o} = 2\left(\frac{k}{\tau_o}\right)\left(\frac{\tau_{sur}a}{l_{sur}}\right)$ describes the ratio of the kinetic and diffusive rates. The term in the first bracket is the rate of step incorporation jumps and in the second is the average rate of the diffusion jumps. Both jumps are over the distance of lattice constant $a$. From that the parameter $\alpha$ is obtained:

$$\alpha = \frac{g}{g + \tanh\left(\frac{y}{l_{sur}}\right)} = \frac{2k\tau_{sur}a}{2k\tau_{sur}a + l_{sur}\tau_o \tanh\left(\frac{y}{l_{sur}}\right)} \tag{6c}$$

A special case of zero supersaturation at the step (BCF boundary condition) is realized in the limit of infinitely fast incorporation, i.e. $k \to \infty$, i.e. $g = \infty$ and $\alpha = 1$ for which the supersaturation distribution given in Eq. 5 is reduced to Eq. 2.

Introduction of the kinetics allows us to describe step asymmetry via different probabilities of the adsorbate incorporation at step: $k_+$ and $k_-$ for the upper and lower terrace respectively. This modifies the basic solution for the terrace only slightly:

$$\sigma = \sigma_v \left[1 - \frac{\alpha \cosh\left(\frac{2z-y}{l_{sur}}\right)}{\cosh\left(\frac{2y}{l_{sur}}\right)} - \frac{\beta \cosh\left(\frac{2z+y}{l_{sur}}\right)}{\cosh\left(\frac{2y}{l_{sur}}\right)}\right] \quad -\frac{y}{2} < z < \frac{y}{2} \tag{7}$$

where the parameters $\alpha$ and $\beta$ determine the lower and higher terrace contributions to the total flux to the step. The condition that $\sigma \geq 0$ limits the possible values of these parameters

$$0 \leq \alpha \leq 1 \tag{8a}$$

$$0 \leq \beta \leq 1 \tag{8c}$$

$$\alpha + \beta \leq \frac{2\cosh\cosh\left(\frac{2y}{l_{sur}}\right)}{1+\cosh\cosh\left(\frac{2y}{l_{sur}}\right)} \tag{8c}$$

The isotropic steps corresponds to $\alpha = \beta$. The solutions given in Eqs 2, 5 and 7 are presented in Fig. 1. From the above diagram it follows that all types of surface kinetics could be recovered by this parameterization, starting from rapid BCF assumption to symmetric

This research was funded in part by National Science Centre, Poland under grant number 2021/41/B/ST5/02764. For the purpose of Open Access, the author has applied a CC-BY-NC-ND public copyright license to any Author Accepted Manuscript (AAM) version arising from this submission.

kinetics, and asymmetric kinetics which could span from BCF fast kinetics to infinite barrier (zero flux). Note that higher value of $\alpha$ and $\beta$ correspond to faster kinetics at higher and lower terrace respectively.

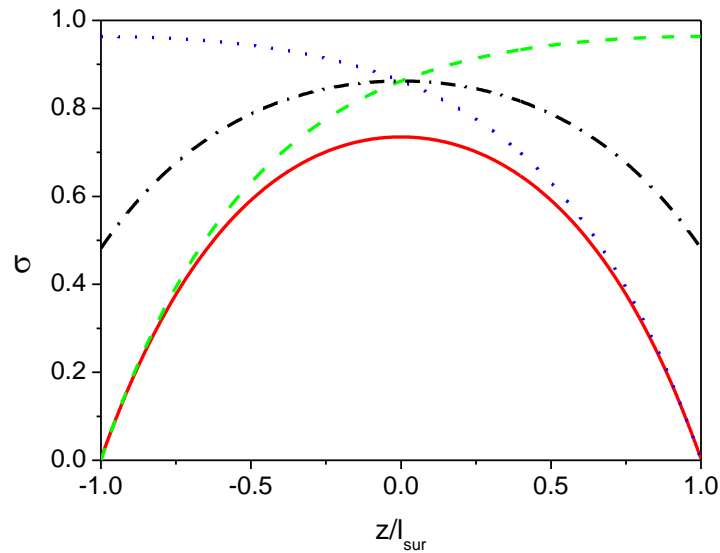

Fig.1. Supersaturation across the terrace for different step kinetics ($y = l_{sur}$): red solid line – infinitely fast (BCF) kinetics ($\sigma = 0$ at the step, $\alpha = 0.9614$ and $\beta = 0.9614$), black dash-dotted line symmetric step barrier kinetics ($\alpha = 0.5$ and $\beta = 0.5$), green dashed line – full asymmetry step kinetics - left ($\alpha = 1.0$ and $\beta = 0.0$),– blue dotted line full asymmetry step kinetics ($\alpha = 0.0$ and $\beta = 1.0$). Steps are at positions -1 and 1 (i.e. at $z = \pm y = \pm l_{sur}$. The selection of the relatively short interstep distance ($y = l_{sur}$) corresponds to strong back force effect related to powerful diffusive interaction of the neighboring steps [43].

These solutions prove that the above expression for the supersaturation across terraces is universal, capable to obtain all physically sound step kinetics. On the other hand this expression allows to express all possible cases, from infinitely rapid symmetric kinetics (BCF solution) through slow symmetric kinetics, to fully asymmetric kinetic with infinitely rapid symmetric kinetics on one side and the complete blockade on the other. Using this representation it is

This research was funded in part by National Science Centre, Poland under grant number 2021/41/B/ST5/02764. For the purpose of Open Access, the author has applied a CC-BY-NC-ND public copyright license to any Author Accepted Manuscript (AAM) version arising from this submission.

possible to find out if the factor can lead to instability of the step system giving rise to multistep surface morphology.

For clarity we assume that upper terrace is on the right and the lower on the left, i.e. the step is moving towards negative z-axis. The conditions of equality of diffusive and incorporation fluxes applies for both terrace edges separately:

- Left side, i.e. lower terrace $\left(z = \frac{y}{2}\right)$

$$\alpha = \frac{g_-}{g_- + \tanh\left(\frac{y}{l_{sur}}\right)} = \frac{2k_-\tau_{sur}a}{2k_-\tau_{sur}a + l_{sur}\tau_o \tanh\left(\frac{y}{l_{sur}}\right)} \quad (9a)$$

- Right side, i.e. upper terrace $\left(z = -\frac{y}{2}\right)$

$$\beta = \frac{g_+}{g_+ + \tanh\left(\frac{y}{l_{sur}}\right)} = \frac{2k_+\tau_{sur}a}{2k_+\tau_{sur}a + l_{sur}\tau_o \tanh\left(\frac{y}{l_{sur}}\right)} \quad (9b)$$

where the side correspond to the side of the step and $g_+ = \frac{2k_+\tau_{sur}a}{l_{sur}\tau_o}$ and $g_- = \frac{2k_-\tau_{sur}a}{l_{sur}\tau_o}$ describe the ratio of the kinetic and diffusive rates. For very fast kinetics, i.e. for $k_+, k_- \to \infty$ we have BCF like distribution, i.e. $\alpha, \beta \to 1$. For slow kinetics, i.e. for $k_+, k_- \to 0$, the parameters are proportional to the kinetic constants, i.e. $\alpha, \beta \to \frac{2k_+\tau_{sur}a}{l_{sur}\tau_o}$, as $\tanh\left(\frac{y}{l_{surf}}\right) \leq 1$. These dependencies may be used to determine the required parameters $\alpha$ and $\beta$. Note that these parameters depend on the terrace width also.

These parameters are used to determine anisotropic step velocity. The effective incorporation rate at the step is:

- Left side, i.e. lower terrace $\left(z = \frac{y}{2}\right)$

$$R_- = j_{dif-}a = \frac{l_{sur}\sigma_v c_{sur-eq} \alpha}{2a\tau_{sur}} \tanh\left(\frac{2y}{l_{sur}}\right) \quad (10a)$$

- Right side, i.e. upper terrace $\left(z = -\frac{y}{2}\right)$

$$R_+ = j_{dif+}a = \frac{l_{sur}\sigma_v c_{sur-eq} \beta}{2a\tau_{sur}} \tanh\left(\frac{2y}{l_{sur}}\right) \quad (10b)$$

From these data the anisotropic step velocity could be obtained as:

$$v = (R_- + R_+)a = \frac{l_{sur}\sigma_v c_{sur-eq}}{2\tau_{sur}}\left[\alpha \tanh\left(\frac{2y_-}{l_{sur}}\right) + \beta \tanh\left(\frac{2y_+}{l_{sur}}\right)\right] \quad (11)$$

where $y_+$ and $y_-$ denote of the upper and lower terrace width, respectively. The derived velocity will be used for stability analysis of the step trains below.

This research was funded in part by National Science Centre, Poland under grant number 2021/41/B/ST5/02764. For the purpose of Open Access, the author has applied a CC-BY-NC-ND public copyright license to any Author Accepted Manuscript (AAM) version arising from this submission.

### III. Stability of parallel equidistant straight steps system.

The stability of the solution describing the stationary motion of parallel BCF step system given by Eq. 2 is analyzed using Mullins-Sekerka variant of Lyapunov approach [60,61]. It is assumed that a small deviation from the basic solution arises by an instantaneous shift of the step $\delta z$, i.e. the step is located at the following position:

$$z = \frac{y}{2} + \delta z = \frac{y}{2} + \delta z_1 \, exp(\lambda t)^{\lambda t} \tag{12}$$

where $\lambda$ is Lyapunov exponent determining the stability of the solution given by Eq. 2. The shift of the step position affects the supersaturation on the neighboring terraces via change of their width that can be expressed by:

$$\sigma(z,t) = \sigma_o(z) + C_1 exp(\lambda t + \gamma z) + C_2 exp(\lambda t - \gamma z) \tag{13}$$

$\sigma_o(z)$ represent solution given by Eq. 2, $C_1$ and $C_2$ are the amplitudes of the disturbance of the supersaturation, that are small compared to $\sigma_o(z)$. Substitution of the solution given by Eq. 13 to Eq. 1a gives the relation between space and time variations:

$$\lambda \tau_{sur} + 1 = \frac{l_{sur}^2 \gamma^2}{4} \tag{14}$$

BCF condition applied to solution given by Eqs. 10 at $z = \frac{y}{2} + \delta z$ and $z = -\frac{y}{2}$ gives the solution $\gamma = 0$ from which the Lyapunov time exponent is obtained via Eq. 11. The time exponent is $\lambda = -\frac{1}{\tau_{sur}}$, i.e. is negative indicating that the parallel step system is stable against deviation of the step position in the asymptotic, Lyapunov linear sense [60,61].

In addition to the analytical approach, the stability of the parallel straight step system may be approached numerically by calculating stepwise time evolution of the parallel step system based on Eq. 3d. The calculated change of the step position at the small time interval $\delta t$ is calculated as $\delta y = v \, \delta t$. At every time interval the velocity of any steps is calculated from Eq. 3d, and the steps position are shifted according to the relation $\delta y = v \, \delta t$.

At the initial time a parallel array of equidistant step is set. The only exception is the central step which is shifted by finite amount $\Delta y$. Naturally this changes the width of the two terraces adjacent to the disturbed step. Thus at the initial time interval $\delta t$ the motion of the three terraces is disturbed: the central one and its two neighbors. Then the more distant step are affected by the terrace width changes.

This research was funded in part by National Science Centre, Poland under grant number 2021/41/B/ST5/02764. For the purpose of Open Access, the author has applied a CC-BY-NC-ND public copyright license to any Author Accepted Manuscript (AAM) version arising from this submission.

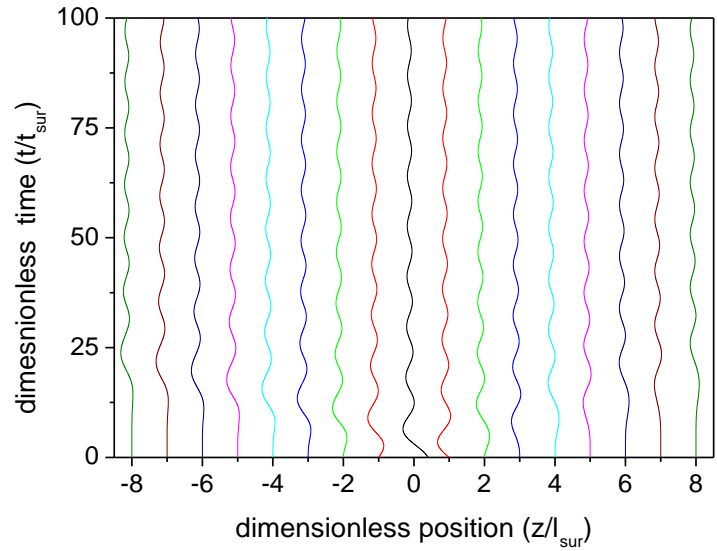

Fig. 2. Time evolution of the step system, consisting of 70 equidistant, parallel straight steps. The steps are straight, parallel during time evolution. The central step is denoted by black color, the symmetric side pairs of the are marked by the same color. Only a central segment of the system containing central 17 steps is shown.

The following units were selected for the simulation: $y = l_{sur}$, $\frac{\Delta y}{l_{sur}} = \frac{\Delta y}{y} = 0.4$, $\sigma_v = 1$, $c_{sur-eq} = 1$. . The boundary condition was zero deviation at both sides, i.e. $\Delta y = 0$. As it is shown, the time evolution of the step system, confirms the results of Lyapunov analysis [60,61]. The central step, which was shifted initially by $\frac{\Delta y}{y} = 0.4$ undergoes oscillations that decay in time . The more distant side steps, that initially have no deviation from equidistant sequence preserve their position for finite time. Afterwards due to the interaction with the neighbors their oscillations are induced. The motion is translated consecutively, spreading the excitation on both sides. Ultimately, even the distant steps oscillations are induced first and afterwards they decay in time, as shown by time evolution of the edge steps. Due to the system size, the oscillations did not reach system boundary during simulation time. Since only a fraction of the step system is displayed, the excitations is finally transferred to the steps located outside the presented area. The amplitude of the deviation decreases in time confirming the overall stability of the parallel equidistant step pattern.

This research was funded in part by National Science Centre, Poland under grant number 2021/41/B/ST5/02764. For the purpose of Open Access, the author has applied a CC-BY-NC-ND public copyright license to any Author Accepted Manuscript (AAM) version arising from this submission.

### IV. Evolution of step train

As it was shown, the single steps in BCF model preserve their spacing even under relatively large deviation. In reality, parallel equidistant steps do not exists. Bulk crystals are finite in size, therefore crystal edges affect the growth. At the edge crystal surfaces is rounded that is equivalent to higher density of the steps which creates natural change of the interstep distance. In addition, it is extremely rare that equidistant step system extends over large distances during growth. Therefore, bulk crystal surfaces consist of the segments having different density of steps. The sequence of the step having zones of much higher step density is known as the step bunch. A variation of the step bunch is the step train where all steps are located at equal distance from each other. The regions surrounding the step bunch or train is characterized by much larger distances between steps. For the purpose of the dynamic analysis, the regions will be denoted as train and inter-train regions. Accordingly the interstep distance in train and in the inter-train space are denoted by $y_{train}$ and $y_{inter}$, respectively ($y_{inter} \gg y_{train}$). As it was argued by several authors, the step train (or step-bunch) is intimately related to the macrostep creation [16,27,28,32]. Following this assumption, in the analysis here we consider a single step train surrounded by the low density areas, which is analytically amenable form of behavior of bunches of the step mentioned in the above references.

This research was funded in part by National Science Centre, Poland under grant number 2021/41/B/ST5/02764. For the purpose of Open Access, the author has applied a CC-BY-NC-ND public copyright license to any Author Accepted Manuscript (AAM) version arising from this submission.

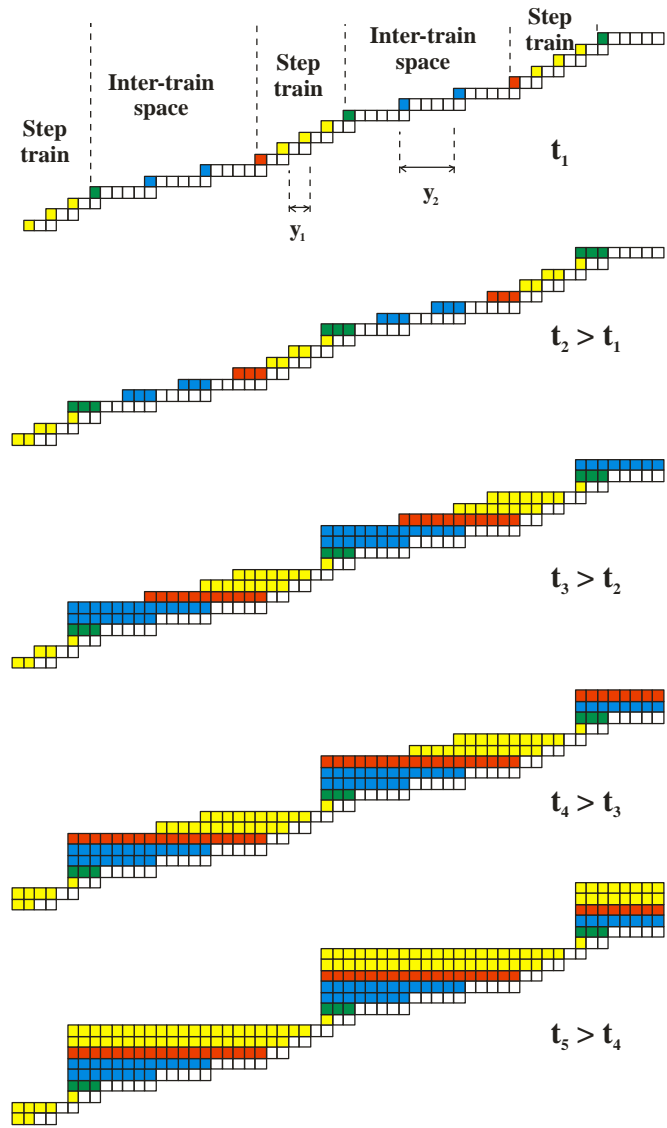

Fig. 3. The time evolution ($t_1 < t_2 < t_3 < t_4 < t_5$) of the nonuniform step system (i.e. system composed of dense step portion (step train) and of sparse step portion (inter-train space) towards macrosteps. The interstep distance in the train is $y_{train}$, in the inter-train spacing is $y_{inter}$, ($y_{train} \ll y_{inter}$). The colors denote different stapes: red – first in the train, green – the last in the train, blue – inside the inter-train space, yellow – inside the train. All colored parts – are the newly deposited layers.

As shown in Fig. 3 we assume that the interstep distance inside and outside (inter) the train and is $y_{train}$ and $y_{inter}$, respectively ($y_{train} \ll y_{inter}$). As the steps interact diffusively only with the nearest neighbors, we consider the two border zones in the front and in the back

This research was funded in part by National Science Centre, Poland under grant number 2021/41/B/ST5/02764. For the purpose of Open Access, the author has applied a CC-BY-NC-ND public copyright license to any Author Accepted Manuscript (AAM) version arising from this submission.

of the step train. The critical problem is the motion of the steps at the zone boundaries: at the front and in the back of the train.

The step at the train front (red in Fig. 3) is surrounded by the two terraces: terrace in front, of the interstep distance $y_2$ (i.e. belonging to the inter-train space) and behind, of the interstep distance $y_1$ (i.e. belonging to the train). Both terraces contribute to the atom incorporation at this step. The steps in the internal part of the train (yellow in Fig. 3) obtain the fluxes from the two terraces of the widths $y_1$. The steps in the inter-trains space (blue in Fig 3) obtain the fluxes from the two terraces of the width $y_2$. For BCF kinetics, the motion in the inter-train space steps is the fastest, the steps inside the train are the slowest [43]. The two steps at the boundaries between the trains and the inter-train space have the intermediate velocity. The front boundary step motion (red in Fig 3), faster than the steps in the train, leads to higher separation inside the train and lower in the preceding inter-train space driving the step system towards uniform spacing in the train and the inter-train space. Therefore, due to motion of the step in the train front no creation of double or multiple steps can occur.

An opposite effect arises at the last step of the train (boundary step), marked by green color in Fig. 3. This step is moving faster than the steps in the train and slower than the steps in the following inter-train space. Therefore the boundary step approaches the preceding steps. That reduces the size of the terrace between the boundary step and the preceding one, slowing both. The size of the terrace in front of the preceding step is increased so it can compensate for the reduction of the terrace behind. In summary the preceding step can proceed with the average velocity of the train. In the simple BCF kinetics the boundary step is potentially capable to catch up with the preceding one provided that the boundary step velocity is higher or equal than the average train velocity. The coalescence occurs when the boundary step is faster than the preceding one in the least suitable case in which the terrace preceding the boundary step disappeared, i.e. when its contribution is zero. The coalescence is therefore attained under the condition that the terrace contribution from behind is sufficient for the boundary step to catch the preceding step. These step velocities, obtained using BCF kinetics, marked "train" – step inside the train and "bound" – boundary step), are:

$$v_{train} = v_\infty tanh\left(\frac{y_{train}}{l_{sur}}\right) \quad (15a)$$

$$v_{bound} = \frac{v_\infty}{2} tanh\left(\frac{y_{inter}}{l_{sur}}\right) \quad (15a)$$

This research was funded in part by National Science Centre, Poland under grant number 2021/41/B/ST5/02764. For the purpose of Open Access, the author has applied a CC-BY-NC-ND public copyright license to any Author Accepted Manuscript (AAM) version arising from this submission.

The factor $1/2$ in the latter case is due to the fact that only single terrace from the following inter-train space (terrace width denoted as $y_{inter}$) contributes to boundary step motion. These dependencies are presented in Fig. 4.

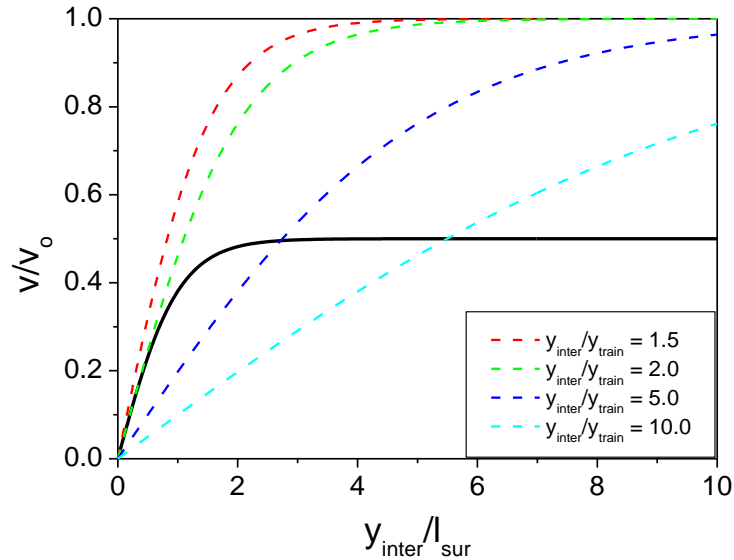

Fig. 4. BCF kinetics relative velocity $v/v_o$, i.e. the ratio of the velocity of the finite size terrace alimented step $v$ to the velocity of single step (i.e. infinite terrace) $v_o$ in function of the inter-train step distance $y_{inter}$ to diffusion length $l_{sur}$ ratio, i.e. ($y_{inter}/l_{sur}$). The lines represent: black solid line - the one side alimented boundary step $v_{bound}$ (Eq. 15a), dashed lines - the two side alimented steps in the train $v_{train}$ (Eq. 15b). The colors mark the inter-train to train terrace width ratio ($y_{inter}/y_{train}$).

As it is shown, for terrace width ratio below 2 (red line in Fig 4), i.e. $y_{inter}/y_{train} \leq 2.0$, the step train is always stable, i.e. the two steps do not merge into single one. For the higher terrace width ratio, the increase of the distance between the steps prevents coalescence. This is due to the fact that the wide terrace center does not contribute proportionally to the step velocity. For low interstep distance (narrow terrace) however, the step coalescence occurs. BCF condition of the step coalescence $v_{free} \geq v_{train}$ can be expressed as

$$\frac{1}{2} tanh\left(\frac{y_{inter}}{l_{sur}}\right) \geq tanh\left(\frac{y_{train}}{l_{sur}}\right) \quad (16)$$

This research was funded in part by National Science Centre, Poland under grant number 2021/41/B/ST5/02764. For the purpose of Open Access, the author has applied a CC-BY-NC-ND public copyright license to any Author Accepted Manuscript (AAM) version arising from this submission.

which can be solved numerically. The plot of this dependence is given in Fig. 5 in which the spacing ratio $\frac{y_{inter}}{y_{train}}$ is plotted in function of step spacing $\frac{y_{inter}}{l_{sur}}$.

$$\frac{y_{inter}}{y_{train}} = 1.95 + 0.324 \left(\frac{0.308 + 0.835\left(\frac{y_{train}}{l_{sur}}\right)}{0.734 - 1.32\left(\frac{y_{train}}{l_{sur}}\right)}\right)^{0.7} \qquad (17)$$

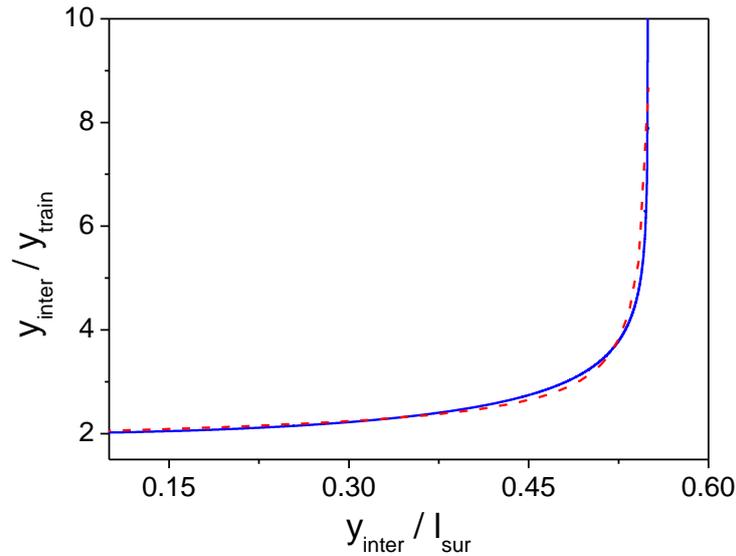

Fig. 5. Lines represent the border between the regions of the step coalescence (above) and of the single step motion (below). The blue line is obtained from solution of Eq. 12, the red line is analytical approximation.

The region above these lines corresponds to the faster velocity of the boundary step and its merger with the preceding one to create a double step. The plot confirms that the single boundary step do not merge, for the $y_{inter}/y_{train}$ ratio below 2 for any $y_{inter}/l_{sur}$ value. In addition it confirms that the increase of the free step spacing is enhancing single step stability, as shown by steep rise of the line marking the merger region. This is especially evident for $y_{inter}/l_{sur} > 0.5$ where single step motion extends to high $y_{inter}/y_{train}$ ratio. Generally the diagram confirms prediction that a single boundary step do not merge for small misorientation of vicinal surfaces. In summary, this diagram proves that single step systems are not transformed into macrosteps in two cases:

i/ relatively uniform steps

This research was funded in part by National Science Centre, Poland under grant number 2021/41/B/ST5/02764. For the purpose of Open Access, the author has applied a CC-BY-NC-ND public copyright license to any Author Accepted Manuscript (AAM) version arising from this submission.

ii/ low misorientation (wide terraces).

Different picture emerges in the case of anisotropic kinetics, as shown in Eq. 11. The step velocities are:

$$v_{train} = (R_- + R_+)a = \frac{v_\infty}{2}\left[\alpha\, tanh\left(\frac{2y_{train}}{l_{sur}}\right) + \beta\, tanh\left(\frac{2y_{train}}{l_{sur}}\right)\right] \quad (18a)$$

$$v_{bound} = R_+ a = \frac{v_\infty}{2} \beta'\, tanh\left(\frac{2y_{inter}}{l_{sur}}\right) \quad (18b)$$

Note that these parameters $(\alpha, \beta)$ depend on the terrace widths, i.e. $\beta \neq \beta'$. The condition of the step coalescence remain valid, i.e. $v_{bound} \geq v_{train}$. These velocities are plotted in Fig. 6.

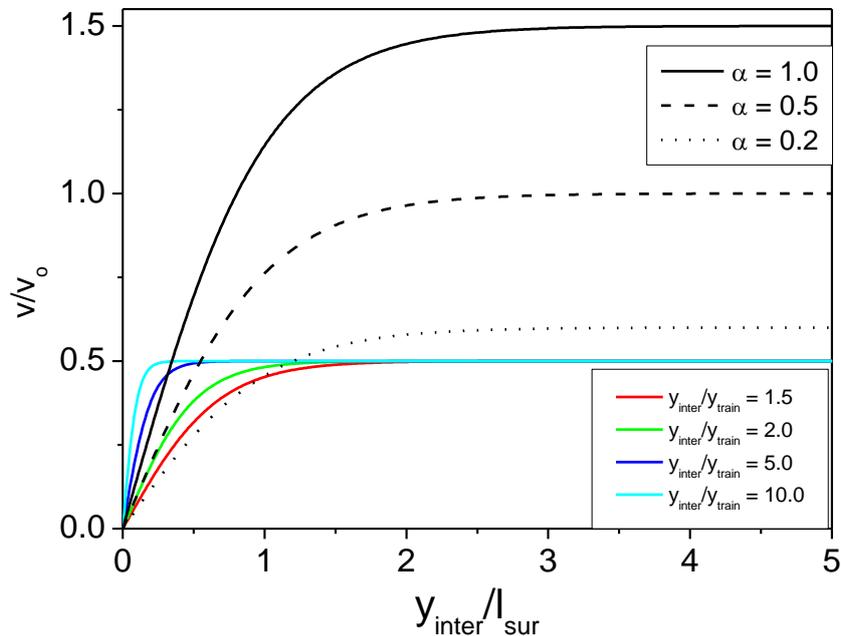

Fig. 6. Relative velocity, i.e. the velocity of the finite size terrace alimented step $v$ divided by the velocity of single step (i.e. infinite terrace) $v_o$ in function of the inter-train step distance $y_{inter}$ to diffusion length $l_{sur}$ ratio. The black represent train steps ( Eq. 18a), for various $\alpha$ values: solid line - $\alpha = 1.0$, dashed line - $\alpha = 0.5$, dotted line - $\alpha = 0.2$. The other lines represent the one side alimented boundary step ( Eq. 18b) marked by the colors representing various the inter-train to train terrace width ratio ($y_{inter}/y_{train}$).

Note that the lines representing boundary step have consistently lower velocity for larger terraces. The result was obtained assuming that the inter-train to terrace width ratio remains

This research was funded in part by National Science Centre, Poland under grant number 2021/41/B/ST5/02764. For the purpose of Open Access, the author has applied a CC-BY-NC-ND public copyright license to any Author Accepted Manuscript (AAM) version arising from this submission.

constant. Thus wider terraces generally favors the single step motion, which dense steps favors coalescence. The condition of the anisotropic step coalescence can be expressed as:

$$\beta' \tanh\left(\frac{2y_{inter}}{l_{sur}}\right) \geq [\alpha + \beta]\tanh\left(\frac{2y_{train}}{l_{sur}}\right) \quad (19)$$

The above relation is different from Eq. 16 which has profound difference on the step coalescence as shown in Fig. 7.

This research was funded in part by National Science Centre, Poland under grant number 2021/41/B/ST5/02764. For the purpose of Open Access, the author has applied a CC-BY-NC-ND public copyright license to any Author Accepted Manuscript (AAM) version arising from this submission.

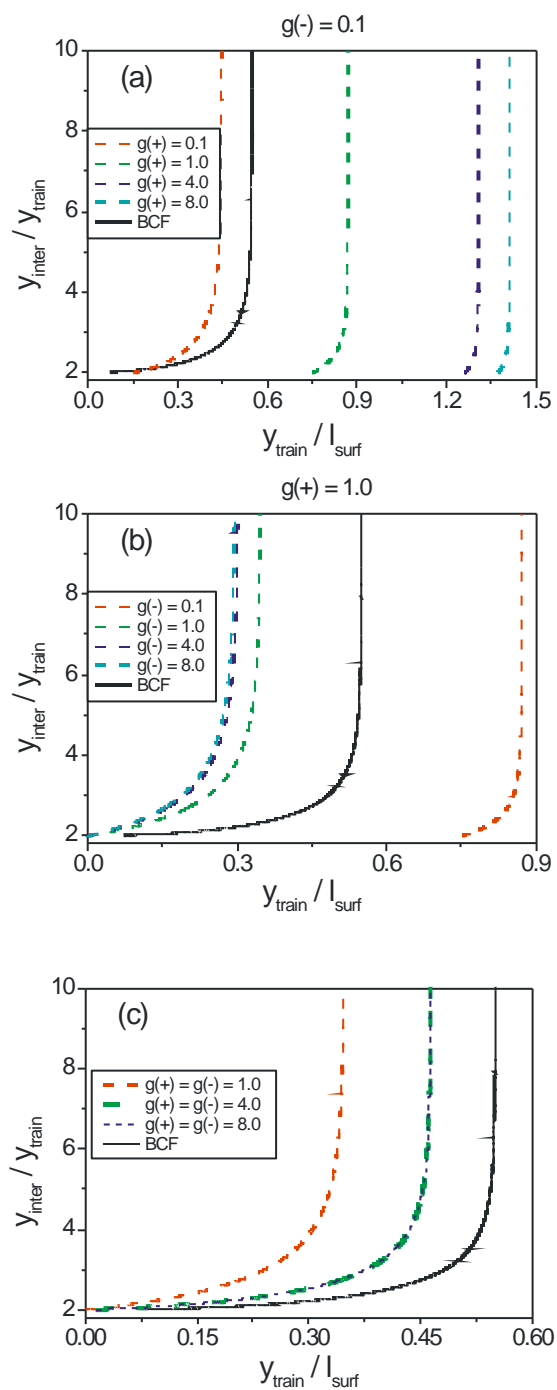

Fig. 7. Lines representing the border between step coalescence region (above) and the region of single step motion (below). The lines are parametrized by the dimensionless ratio of kinetic

This research was funded in part by National Science Centre, Poland under grant number 2021/41/B/ST5/02764. For the purpose of Open Access, the author has applied a CC-BY-NC-ND public copyright license to any Author Accepted Manuscript (AAM) version arising from this submission.

and diffusive rates $g_+ = \frac{2k_+\tau_{sur}a}{l_{sur}\tau_o}$ and $g_- = \frac{2k_-\tau_{sur}a}{l_{sur}\tau_o}$ from upper and lower terrace, respectively: (a) $g_- = 0.1 = const$, (b) $g_+ = 1.0 = const$, (c) $g_+ = g_-$.

As it is shown the single step is stable against coalescence for $y_{inter}/y_{train} \leq 2.0$, irrespective of the values of other factors, i.e. kinetic coefficients and $y_{inter}/l_{sur}$ ratio. Wide terraces favors the single step structure stability against coalescence. The asymmetric step merger factors may be summarized as follows:

i/ faster upper terrace kinetics, i.e. increase of $g_+$, favors merger

ii/ faster lower terrace kinetics, i.e. increase of $g_-$, favors single step motion

iii/ simultaneous increase of the kinetics of both terraces favors merger

iv/ BCF is more prone to step merger than any finite kinetic coefficient, equal for both sides, i.e. $k_- = k_+ = k$.

For small interstep distances, the step kinetics becomes all important, as the transport occur at small distance only. This may be analyzed using the right to left side kinetics contribution to the step dynamics, expressed as

$$\frac{\alpha}{\beta} = \frac{k_-}{k_+}\frac{\left[2k_+\tau_{sur}a + l_{sur}\tau_o \tanh\left(\frac{2y_{train}}{l_{sur}}\right)\right]}{\left[2k_-\tau_{sur}a + l_{sur}\tau_o \tanh\left(\frac{2y_{inter}}{l_{sur}}\right)\right]} \cong \frac{k_-}{k_+}\frac{[k_+\tau_{sur}a + \tau_o y_{train}]}{[k_-\tau_{sur}a + \tau_o y_{inter}]} \tag{20}$$

The condition for single step coalescence is

$$\frac{y_{inter}}{y_{train}} \geq \frac{\alpha+\beta}{\beta'} \approx 1 + \frac{k_-}{k_+} \tag{21}$$

Thus in case of extremely fast kinetics, the condition for coalescence is identical to BCF data. In the slow kinetics regime the coalescence is favored by the faster incorporation from lower terrace as expected.

The presented scenario of the train evolution leads to two different outcomes depending on the coalescence propensity:

- Advance of the step in the front of the train leads to train dilution, finally creating uniform vicinal surface
- Advance of the step in the rear of the train potentially leads to creation of macrostep/superterrace structure. The multistep is created by the creation of a double step to which the following single steps are attached. The condition of the step merger were defined above.

This research was funded in part by National Science Centre, Poland under grant number 2021/41/B/ST5/02764. For the purpose of Open Access, the author has applied a CC-BY-NC-ND public copyright license to any Author Accepted Manuscript (AAM) version arising from this submission.

It is worth to note that the presented model provides explanation of basic mechanism of the macrostep creation. As shown it is induced by the difference of step density at the surface which may in turn be caused by plethora of well identified factors. Among them may be differences is surface preparation, the kink poisoning, the growth obstacles, including the foreign phase inclusions attached to the surface, the extended defects at the surface or finally the different flux from the vapor affecting step trains, as shown by Chernov in Ref 62. In principle, the step meandering may to some degree affect the density of step and therefore contribute to creation of macrosteps [57,63-65]. These factors will not be be investigated in this paper.

### V. Macrostep and superterraces - growth mode

The above presented macrostep creation scenario explains the existence of the surface structure that is essential for the growth of bulk crystals. Nevertheless, growth of the bulk crystals cannot proceed exclusively via macrostep motion because macrostep translation requires accumulation of huge amount of material. In order this to occur such accumulation should be accompanied by the absence of single step emission from the macrostep bottom. As there is not large energy barrier against single step emission from the macrostep bottom, the propagation of the single steps across the superterraces is the dominant growth mode of the bulk crystals. Due to this propagation, the surface consists of the macrosteps and the segments of vicinal surfaces, i.e. separated single steps. These vicinal segments will be called superterraces. In short, the surface consist of macrosteps and superterraces which are vicinal segments. The single steps are emitted from the macrostep, move across vicinal surface and arrive and next macrosteps. This is the dominant growth mode. Macrostep motion as such plays relatively inferior role, but it could affect the growth and properties of the grown crystals locally. Thus the growth proceeds by emission of the single steps, their flow across superterraces and their arrival at the neighboring macrosteps.

The growth is therefore controlled by the emission of single steps from the macrostep bottom. Thus the relevant question is to find the step emission waiting period. From this the frequency of the single step emission, their separation and other characteristics could be obtained. The waiting period starts immediately after the preceding step was emitted. During that period, the material is incorporated at the step. The first preceding step advances across the terrace, therefore the position changes in time, denoted as $z_1(t)$. The step is emitted when the

This research was funded in part by National Science Centre, Poland under grant number 2021/41/B/ST5/02764. For the purpose of Open Access, the author has applied a CC-BY-NC-ND public copyright license to any Author Accepted Manuscript (AAM) version arising from this submission.

average accumulation of the atoms at the single site is 1, i.e. in average single atom is attached at along the whole step at each site. Therefore the single step emission time $t_1$ is determined from the relation

$$1 = \int_0^{t_1} dt\, R_-(z_1(t)) = \int_0^{t_1} dt \left( \frac{l_{sur} \sigma_v c_{sur-eq}\, \alpha}{2a\tau_{sur}} tanh\left(\frac{z_1(t)}{l_{sur}}\right) \right) \quad (22)$$

where the terrace width is $z_1(t) = 2y(t)$. Therefore the position of the step 1 is:

$$z_1(t) = \int_0^t dt\, v_1(t) = \frac{l_{sur} \sigma_v c_{sur-eq}}{2\tau_{sur}} \int_0^t dt \left[ \alpha_1 tanh\left(\frac{z_2(t)-z_1(t)}{l_{sur}}\right) + \beta_1 tanh\left(\frac{z_1(t)}{l_{sur}}\right) \right] \quad (23)$$

where by $z_2(t)$ we denote the position of the second step. The position that the first step attain at the time $t_1$ is $Z_1 = z_1(t_1)$. At the time $t = 0$, the position of the step 2 is $Z_1$, i.e. $z_2(0) = Z_1$. The motion of the step 2 depends on the position of the step 1 and also on the position of step 3, denoted by $z_3(t)$:

$$z_2(t) = Z_1 + \int_0^t dt\, v_2(t) = \frac{l_{sur} \sigma_v c_{sur-eq}}{2\tau_{sur}} \int_0^t dt \left[ \alpha_2 tanh\left(\frac{z_3(t)-z_2(t)}{l_{sur}}\right) + \beta_2 tanh\left(\frac{z_2(t)-z_1(t)}{l_{sur}}\right) \right]$$
$$(24)$$

The position of the step 2 at the time $t_1$ is $Z_2$. Naturally, the motion of step 3 is affected by the following step, i.e. by step 2. This leads to variation of the step velocity in time, and it could also lead to the oscillations, such as shown in Fig. 2. The motion of more distant step is only weakly disturbed, therefore in order to simplify the problem it is assumed that step 3 moves with the constant velocity:

$$z_3(t) = Z_2 + v_3\, t \quad (25)$$

In order to determine the final velocity it is assumed that this velocity is given by

$$v_3 = \frac{l_{sur} \sigma_v c_{sur-eq}}{2\tau_{sur}} [\alpha_3 + \beta_3]\, tanh\left(\frac{\Delta z_3}{l_{sur}}\right) \quad (26)$$

The parameters $\alpha_i$ and $\beta_i$ are expressed using Eqs 9 and employing appropriate thickness of the terraces. The solution of these equation set is cumbersome therefore the approximate solution will be used. The step is detached from the surface and gradually accelerates to constant velocity. This could be approximated by the following function:

$$z(t) = v_3\, t\, tanh\left(\frac{t}{t_1}\right) \quad (27)$$

This function has correct, i.e. approximately quadratic dependence for small t and linear for large t. Therefore step moves with velocity $v_3$ at large t, i.e. at far distance from the macrostep.

This research was funded in part by National Science Centre, Poland under grant number 2021/41/B/ST5/02764. For the purpose of Open Access, the author has applied a CC-BY-NC-ND public copyright license to any Author Accepted Manuscript (AAM) version arising from this submission.

The transition to linear dependence occurs approximately at the time $2t_1$. this step moves with the velocity $v_3$, equal to the velocity of the average step at the superterrace. The steps at superterrace are separated by $\Delta z_3 = v_3 t_1$ which leads to the equation of motion

$$v_3 = \frac{l_{sur}\sigma_v c_{sur-eq}}{2\tau_{sur}}[\alpha_3 + \beta_3]tanh\left(\frac{v_3 t_1}{l_{sur}}\right) \quad (28)$$

Since the step is alimented from both sides, the difference in kinetics could be neglected, thus $\alpha_3 \approx \beta_3 = \frac{g}{g+tanh\left(\frac{v_3 t_1}{l_{sur}}\right)}$ where $g = \frac{\tau_{sur} k a}{l_{sur}}$ is calculated using arithmetic mean value of both kinetic coefficient $k = \frac{k_+ + k_-}{2}$. The distance between steps $\Delta z_3$ is equal to the saturation velocity of single step far from macrostep $v_3$ and the time of the emission of the new step at the $t_1$., i.e. $\Delta z_3 = v_3 t_1$. This allows us to obtain the dimensionless interstep distance by division by surface diffusion length $l_{sur}$, i.e. $\kappa \equiv \frac{\Delta z_3}{l_{sur}} = \frac{v_3 t_1}{l_{sur}}$. The equation for $\kappa$ is

$$\kappa[g + tanh(\kappa)] = \sigma_v c_{sur-eq}\left(\frac{t_1}{\tau_{sur}}\right) g\, tanh(\kappa) \quad (29)$$

which can be solved, as shown in Fig. 8.

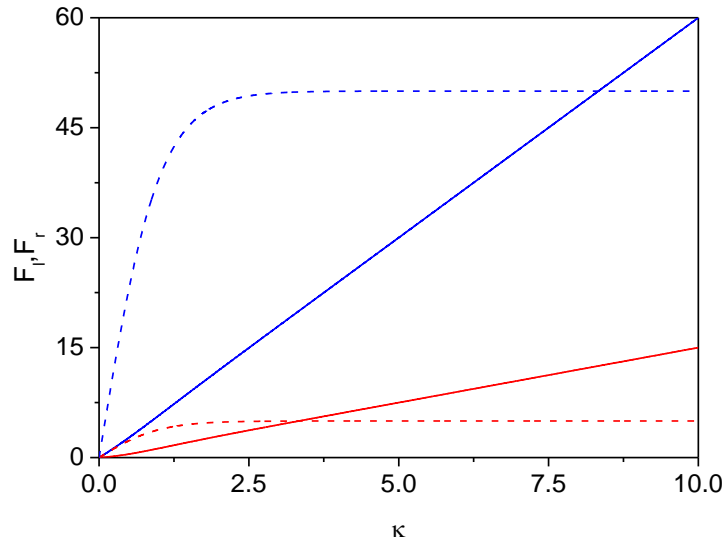

Fig.8. Plots of the sides of Eq. 29: left side, i.e. $F_l = \kappa[g + tanh(\kappa)]$ and right side, i.e. $F_r = \sigma_v c_{sur-eq}\left(\frac{t_1}{\tau_{sur}}\right) g\, tanh(\kappa)$ in function of dimensionless distance $\kappa$. $F_l$ and $F_r$ are represented by solid and dashed lines, respectively. The red and blue color denote relatively slow ($g = 0.5$)

This research was funded in part by National Science Centre, Poland under grant number 2021/41/B/ST5/02764. For the purpose of Open Access, the author has applied a CC-BY-NC-ND public copyright license to any Author Accepted Manuscript (AAM) version arising from this submission.

and relatively fast ($g = 5$) surface kinetics, respectively. The time scaling factor was $\sigma_v c_{sur-eq} \left( \frac{t_1}{\tau_{sur}} \right) = 10$.

The small $\kappa$ dependence limits the existence of the solution of Eq. 29. The two regimes of slow and fast kinetics have different solution existence criteria and values:

i/ slow kinetics ($g < 1.0$)

$$\frac{\sigma_v c_{sur-eq} t_1 \, ka}{l_{sur}} > 1 \quad (30a)$$

$$v_3 = \frac{\sigma_v c_{sur-eq} l_{sur}}{\tau_{sur}} \quad (30b)$$

ii/ fast kinetics ($g > 1.0$)

$$\frac{\sigma_v c_{sur-eq} t_1}{\tau_{sur}} > 1 \quad (31a)$$

$$v_3 = \sigma_v c_{sur-eq} ka \quad (31b)$$

This determines the velocity of the steps across the superterrace, emitted from the macrostep.

It also useful to find the emission time $t_1$ using Eq. 22 employing the approximate solution given by Eq. 27. From these equations the following relation may be derived:

$$t_1 = \tau_{sur} \left( \frac{a}{l_{sur}} \right) \frac{2}{\sigma_v c_{sur-eq} \, I(\kappa, g_-)} \quad (32a)$$

where the integral $I(\kappa, g_-)$ is given by

$$I(\kappa, g_-) = \int_0^1 d\varphi \, \frac{g_- \tanh(\kappa \varphi \tanh(\varphi))}{g_- + \tanh(\kappa \varphi \tanh(\varphi))} \quad (32b)$$

The integral may be easily estimated as $0 \leq I(\kappa, g_-) \leq 1$. From this the approximate value of the emission time $t_1$ as

$$t_1 \approx \tau_{sur} \left( \frac{a}{l_{sur}} \right) \frac{2}{\sigma_v c_{sur-eq}} \quad (33)$$

As expected, lower equilibrium density and supersaturation in the vapor, i.e. lower maximal coverage of the surfaces entails longer emission time. This could be used to determine the step separation in the superterrace, i.e. terrace width as

$$\Delta z_3 = v_3 \, t_1 \quad (34)$$

where the values are given by Eqs 30, 31, 32 and 33.

Note that the natural factor contributing to the step trains creation is the presence of the crystal edge. In this case the crystal surface is very prone to macrostep creation as the flat portion is behind, most suitable arrangement for the step creation. It is therefore expected that

This research was funded in part by National Science Centre, Poland under grant number 2021/41/B/ST5/02764. For the purpose of Open Access, the author has applied a CC-BY-NC-ND public copyright license to any Author Accepted Manuscript (AAM) version arising from this submission.

the macrosteps are located close to crystal edges. It is worth noting that in such a case, the superterraces may be not wide enough to accommodate many terraces. Therefore the single step emission is slowed down preserving a large number of macrosteps located in close distance at the crystal edges.

An additional note has to be added regarding to macrostep stability and time evolution. As it was explained, the macrostep is not a result of creation of the structure which is stable by some additional mechanism and therefore stationary, i.e. time independent. The macrostep persist in time, depending on the balance between accumulation and emission of the single steps. It exists in the macrostep ensemble, where the single steps travel between macrosteps across the superterraces. The distance between the single steps on the terrace may be obtained by division of superterrace width by the average distance between steps $\Delta z_3$, given by Eqs. 30-34. Depending on the accumulation and emission the step can persist for long time.

### VI. Creation of overhangs and inclusions

The above outlined scenario describes creation and activity of the macrosteps during growth. It is not complete. The macrosteps may potentially create inclusions, i.e. incorporation of the volumetric bulk of the other phase. The inclusion is created by creation of the overhangs at the macrostep edge and their build-up and final coalescence with the surface. The volume underneath is filled by mother phase or vacuum.

The creation of the overhang proceeds via creation of stable atomic configuration on top of the macrostep by combination of the surface diffusion motion and impingement from the vapor. This is illustrated in Fig. 9 where the basic atomic processes leading to overhang instability of the macrostep are illustrated.

This research was funded in part by National Science Centre, Poland under grant number 2021/41/B/ST5/02764. For the purpose of Open Access, the author has applied a CC-BY-NC-ND public copyright license to any Author Accepted Manuscript (AAM) version arising from this submission.

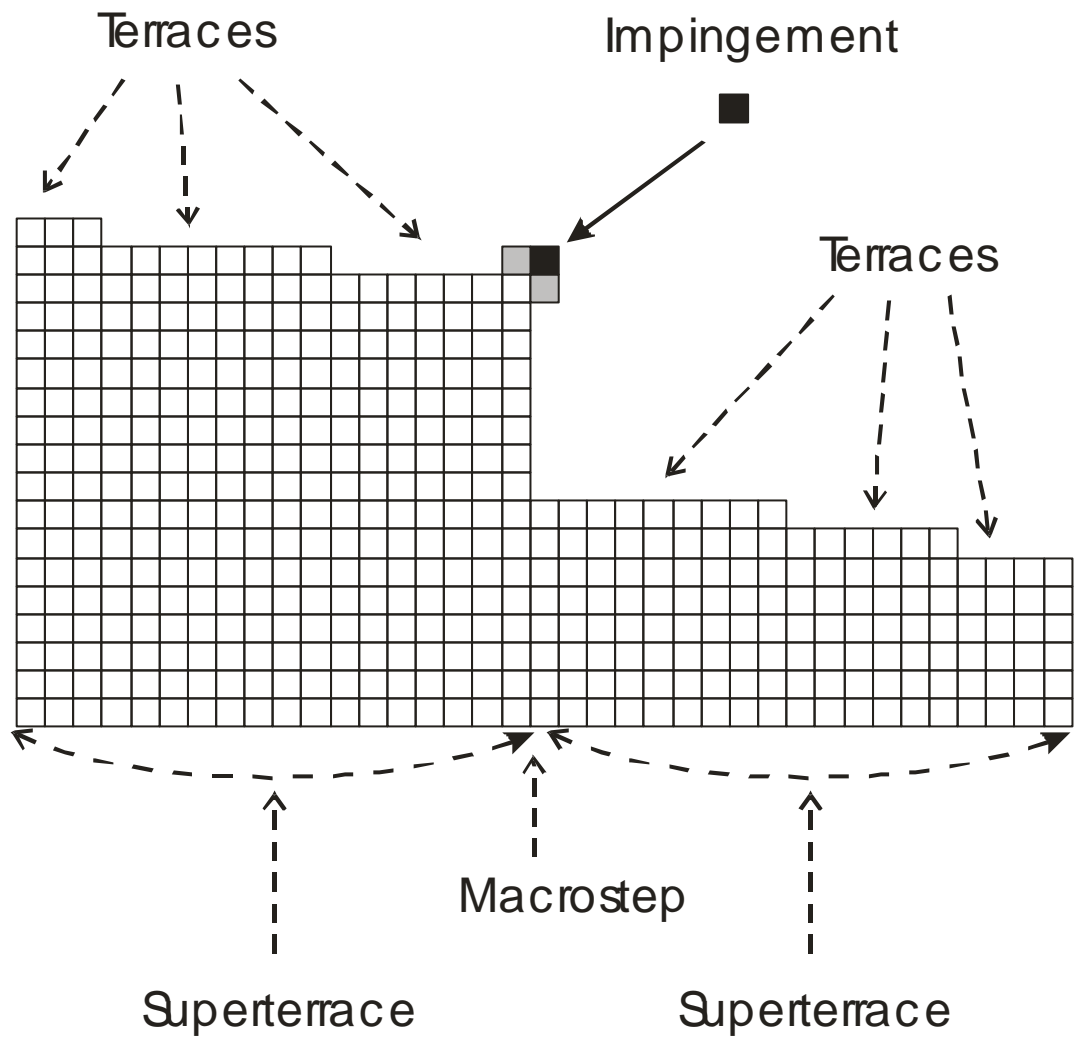

Fig. 9. Basic geometric constructs and the atomistic processes leading to stable atomic configuration giving rise to overhang instability of the macrostep (side view). White boxes represent atom in the solid lattice. Gray boxes represent atoms arriving at the specified position due to surface diffusion jumps and also by impingement from the vapor. Black box represents the atom impinging from the vapor.

The basic principle of instability emergence was already proposed by the authors some time ago [68,69]. According to this formulation, the instability starts by creation of energetically stable atomic configuration at the crystallization front. This could be determined by time of the creation of the stable atomic configuration $\tau_{ovh}$ as shown in Fig. 9. We assume that the probability of the occupation of the adjacent surface sites is equal to the occupation of the

This research was funded in part by National Science Centre, Poland under grant number 2021/41/B/ST5/02764. For the purpose of Open Access, the author has applied a CC-BY-NC-ND public copyright license to any Author Accepted Manuscript (AAM) version arising from this submission.

surface sites, $p = c_{sur} = c_{sur-eq}\sigma$. The surface site vapor impingement frequency is $v_v = \tau_o^{-1} exp\left(-\frac{3\phi}{kT}\right)$, that leads to

$$\tau_{ovh} = \frac{2}{3v_v p^2} = \frac{2\tau_o}{3c^2} exp\left(-\frac{3\phi}{kT}\right) \qquad (34)$$

where the impingement frequency of the corner site was assumed to be larger by a factor of $3/2$ that this frequency of the site at flat surface. The difference is due to larger corner of exposure.

The following evolution of the overhang is related to complex vapor bulk and surface diffusion interplay which depends also on the ballistic and diffusive transport mode in the vapor. As it was shown this leads to corner instability preferential growth for the following length ratio $L \geq 4\lambda$. In the presented case for the size of the growing object $L$ the macrostep height $H$ should be selected, $L = H$. The mean free path in the vapor $\lambda$ is typically much larger than the macrostep height $H$, therefore for the vapor growth the factor $3/2$ should be omitted. In case of the growth from the liquid phase $\lambda < a \ll H$, therefore the factor should be used.

### VII. Summary

The analysis of the crystal growth at the vicinal surfaces was made using various surface kinetics models. It was shown that the supersaturation at the terraces could be represented by the same family of solutions. The values of the solution parameters were derived in function of the atom incorporation at the steps kinetic coefficients.

The presented results confirmed the earlier reports that linear Lyapunov type analysis predicts stability of the equidistant step systems for isotropic, rapid kinetics such as described by BCF. The time integration of the motion of such a system shows more complex behavior, in which the localized disturbance is spreading in both sides. Ultimately, this behavior is in agreement with the linear stability analysis results showing that the amplitude of the deviation of the step position decreases in time.

The new, completely different scenario of the instability leading to creation of the macrostep was proposed. According to this scenario the creation of macrosteps is related to surface diffusion kinetics only. The macrosteps are created due to differences in the step densities at the vicinal surface. The high step density regions, i.e. step trains are surrounded by low step density (inter-train) segments . The critical is the dynamics of the steps at the train

This research was funded in part by National Science Centre, Poland under grant number 2021/41/B/ST5/02764. For the purpose of Open Access, the author has applied a CC-BY-NC-ND public copyright license to any Author Accepted Manuscript (AAM) version arising from this submission.

edges. The step in the train front accelerates due to wider terrace in front that leads to the decrease of the step density in the train and eventually to the equidistant pattern.

The boundary step in the train rear also accelerates due to more abundant alimentation by the wider terrace behind. This step approaches the preceding step. This slows their motion, both the boundary step and the preceding one. Most drastically the terrace between boundary step and the preceding one disappears, providing no alimentation to their motion. If the boundary step alimentation by the terrace behind induced faster motion than the step train the boundary step merge with the preceding one leading to creation of a double step. The condition for step coalescence was derived. The step cannot catch the previous one if the step density ratio in and out of the train is lower than 2. For higher step density ratio the condition derived indicates that low density of the steps promotes single step motion thus preventing step coalescence. The fast incorporation from the upper terrace strongly favors step coalescence. The fast incorporation from the lower terrace favors single step motion.

The creation of double step creates structure that moves slower as it needs twice as large alimentation to move. Thus the double step is relatively slow allowing the following step to accelerate and catch creating possible triple step. Thus the steps behind catch up creating multistep structure. Finally, it can accumulate many steps to create macrostep.

The macrosteps are not basically mobile as they are alimented from the terraces and the alimentation needed to move is very large. Moreover, only upper terrace alimentation leads to step motion. The alimentation from the terrace at the macrostep bottom leads to emission of a single step which moves forward. Therefore the final surface structure consists of macrosteps and superterraces. At the superterraces the number of single steps are present, i.e. the superterraces are vicinal surface segments. The single steps are moving across superterraces toward the preceding macrosteps. Thus the single step motion is dominant crystal growth mode despite the presence of the macrosteps. The frequency of the single step emission and the velocity of its motion was derived.

The macrostep could be prone to creation of the overhang which is due to the surface dynamics coupling to impingement from the mother phase. The promoting factor of the instability is the angular preferential access of the bulk material to the macrostep edge, leading to diffusive instability. Therefore it is expected that harmful influence of the macrosteps leading to creation of inclusions and dislocation is stronger during growth from the liquid phase.

This research was funded in part by National Science Centre, Poland under grant number 2021/41/B/ST5/02764. For the purpose of Open Access, the author has applied a CC-BY-NC-ND public copyright license to any Author Accepted Manuscript (AAM) version arising from this submission.

**ACKNOWLEDGMENT.**

The research was partially supported by [National Science Centre, Poland] grants number 2016/23/B/ST5/02278 and 2021/41/B/ST5/02764. Theoretical calculations were carried out with the support of the [Interdisciplinary Centre for Mathematical and Computational Modelling (ICM) University of Warsaw] under computational allocations no [G83-19] [GB76-25] and [GB84-23].

This research was funded in part by National Science Centre, Poland under grant number 2021/41/B/ST5/02764. For the purpose of Open Access, the author has applied a CC-BY-NC-ND public copyright license to any Author Accepted Manuscript (AAM) version arising from this submission.

This research was funded in part by National Science Centre, Poland under grant number 2021/41/B/ST5/02764. For the purpose of Open Access, the author has applied a CC-BY-NC-ND public copyright license to any Author Accepted Manuscript (AAM) version arising from this submission.

This research was funded in part by National Science Centre, Poland under grant number 2021/41/B/ST5/02764. For the purpose of Open Access, the author has applied a CC-BY-NC-ND public copyright license to any Author Accepted Manuscript (AAM) version arising from this submission.

This research was funded in part by National Science Centre, Poland under grant number 2021/41/B/ST5/02764. For the purpose of Open Access, the author has applied a CC-BY-NC-ND public copyright license to any Author Accepted Manuscript (AAM) version arising from this submission.

This research was funded in part by National Science Centre, Poland under grant number 2021/41/B/ST5/02764. For the purpose of Open Access, the author has applied a CC-BY-NC-ND public copyright license to any Author Accepted Manuscript (AAM) version arising from this submission.

This research was funded in part by National Science Centre, Poland under grant number 2021/41/B/ST5/02764. For the purpose of Open Access, the author has applied a CC-BY-NC-ND public copyright license to any Author Accepted Manuscript (AAM) version arising from this submission.